\documentclass[12pt]{iopart}

\usepackage{graphicx}

\begin{document}

\title[Stability analysis of dynamic thin shells]
{Stability analysis of dynamic thin shells}

\author{Francisco S. N. Lobo\footnote[1]{flobo@cosmo.fis.fc.ul.pt}
 and Paulo Crawford\footnote[2]{crawford@cosmo.fis.fc.ul.pt} }

\address{Centro de Astronomia
e Astrof\'{\i}sica da Universidade de Lisboa,\\
Campo Grande, Ed. C8 1749-016 Lisboa, Portugal}

%%%%%%%%%%%%%%%%%%%%%%%%%%%%%%%%%%%%%%%%%%%%%%%%%%%%%%%%%%%%%%%%%

\begin{abstract}

We analyze the stability of generic spherically symmetric thin
shells to linearized perturbations around static solutions. We
include the momentum flux term in the conservation identity,
deduced from the ``ADM'' constraint and the Lanczos equations.
Following the Ishak-Lake analysis, we deduce a master equation
which dictates the stable equilibrium configurations. Considering
the transparency condition, we study the stability of thin shells
around black holes, showing that our analysis is in agreement with
previous results. Applying the analysis to traversable wormhole
geometries, by considering specific choices for the form function,
we deduce stability regions, and find that the latter may be
significantly increased by considering appropriate choices for the
redshift function.

\end{abstract}


\maketitle

%%%%%%%%%%%%%%%%%%%%%%%%%%%%%%%%%%%%%%%%%%%%%%%%%%%%%%%%%%%%%%%%
%%%%%%           Introduction                       %%%%%%%%%%%%
%%%%%%%%%%%%%%%%%%%%%%%%%%%%%%%%%%%%%%%%%%%%%%%%%%%%%%%%%%%%%%%%

\section{Introduction}

The study of hypersurfaces of discontinuity plays a fundamental
role in general relativity. Pioneering work can be traced back to
Sen \cite{Sen}, Lanczos \cite{Lanczos}, Darmois \cite{Darmois},
and later by O'Brien and Synge~\cite{Synge},
Lichnerowicz~\cite{Lich}, and Israel \cite{Israel}, amongst
others. Several approaches to deduce the geometric conditions of
the surface layer were carried out. For instance, Sen in a
relatively unknown paper \cite{Sen}, used continuous
four-dimensional coordinates across the layer, an approach later
applied by Lichnerowicz, and O'Brien and Synge. Darmois expressed
the surface properties of the discontinuity of the extrinsic
curvature across the layer as a function of the surfaces intrinsic
coordinates, which was later generalized by Israel \cite{Israel},
using independently defined four-dimensional coordinate systems
$x^{\mu}_\pm$ of the two different manifolds glued at the junction
surface.
Since then, the hypersurfaces of discontinuity have had an
extensive range of applicability. It is perhaps important to
emphasize that the matching of an interior dust solution to an
exterior Schwarzschild spacetime, by Oppenheimer and Snyder,
provided the first insights of the gravitational collapse into a
black hole \cite{Opp}. One can mention more recent applications of
the thin shell formalism, namely, gravitational collapse of
radiating shells, the evolution of bubbles and domain walls in
cosmological settings and in inflationary models, the structure
and dynamics of voids in the large scale structure of the
universe, shells around black holes and their respective
stability, signature changes, matchings of cosmological solutions,
and applications to the Randall-Sundrum brane world scenario to
wormhole physics. An analysis in thick gravitating walls expanded
in powers of the thickness of the wall has also been carried out
\cite{Garfinkle,KM}. The lightlike limit was analyzed in
Ref.~\cite{BI}, and an equivalence between the Darmois-Israel
formalism and the distributional method for thin shells was
established in Ref. \cite{Mansouri}. Due to the extensive
applications of the Darmois-Israel formalism, a computer algebra
system was implemented to aid relativists in the evaluation of
junction conditions and the parameters associated with thin
shells~\cite{Musgrave}.

The application of the Darmois-Israel formalism to black holes and
wormhole physics is of a particular interest. Relatively to black
holes, in an interesting paper \cite{FMM}, the Schwarzschild
metric inside the event horizon was matched to the de Sitter
solution at a spacelike junction surface. The respective
properties of the layer were explored, and it was found that
instead of a singularity, a closed world can be formed inside the
black hole. In Ref. \cite{FHK}, a thin shell was constructed
around a black hole and its characteristics were explored by
imposing the energy conditions, and in Ref. \cite{BLP} the
respective stability of the thin shell was analyzed against
spherically symmetric perturbations about a static solution.
Relatively to wormhole physics, using the cut-and-paste technique,
Visser constructed thin-shell wormholes, and considered a partial
stability analysis by imposing specific equations of
state~\cite{VisserPRD,VisserNP,VisserPLB,Visser}. In Ref.
\cite{Poisson} thin-shell Schwarzschild wormholes were considered
and the respective stability to spherically symmetric
perturbations around static solutions was analyzed, in the spirit
of \cite{BLP}. It was later found that the inclusion of a charge
\cite{Eiroa} and of a positive cosmological constant
\cite{Lobolinear} significantly increases the stability regions.
Specific wormhole solutions were also constructed by matching an
interior traversable wormhole solution to exterior vacuum
spacetimes at a junction interface
\cite{LLQ,Lobo,Lobo-CQG,LL-PRD}. Thin shells with a zero surface
energy density were analyzed in Ref. \cite{LLQ}, dust shells in
Ref. \cite{Lobo}, generic surface stresses in Ref.
\cite{Lobo-CQG}, and a similar analysis for plane symmetric
traversable wormholes in an anti-de Sitter background was
extensively studied in Ref. \cite{LL-PRD}.

Ishak and Lake developed a formalism to analyze the stability of
spherically symmetric thin shells by imposing spacetimes that
satisfy the transparency condition \cite{Ishak}, which amounts to
considering solutions that do not contribute with the
discontinuity flux term in the conservation identity. The dynamics
of timelike spherical thin shells, satisfying the transparency
condition were also analyzed in Ref. \cite{Goncalves}. In this
work, we include this momentum flux term, which severely
complicates the analysis, and following the Ishak-Lake approach,
we analyze the stability of generic spherically symmetric thin
shells to linearized perturbations about static equilibrium
solutions. We then deduce a master equation dictating the
stability regions. Considering the transparency condition, we
study the stability of thin shells around black holes, showing
that our analysis is in agreement with previous results. We also
apply the general formalism to specific wormhole solutions, and by
considering particular form functions, we verify that the
stability regions can be significantly increased by taking into
account appropriate choices of the redshift function. In this
context, a specific application of the general formalism developed
in this paper has been applied to traversable wormhole solutions
\cite{dyn-phantomWH} supported by phantom energy
\cite{Sushkov,Lobo-phantom}.

The plan of this paper is as follows: In Section II, we outline a
brief summary of the Darmois-Israel formalism, which shall be used
throughout the work. In Section III, we present two generic
spherically symmetric spacetimes matched together at a junction
interface and deduce the surface stresses of the thin shell. Based
on the inclusion of the momentum flux term in the conservation
identity, fundamental relationships for analyzing the stability
regions are also deduced. From the equation of motion, a master
equation, dictating the stable equilibrium configurations, is
deduced. In Section IV, we shall apply the formalism developed to
thin shells around black holes. In Section V, the formalism is
applied to traversable wormholes, imposing several choices for the
form and redshift functions. Finally, in Section VI, we conclude.

\section{The Darmois-Israel formalism}

Consider two distinct spacetime manifolds, ${\cal M_+}$ and ${\cal
M_-}$, with metrics given by $g_{\mu \nu}^+(x^{\mu}_+)$ and
$g_{\mu \nu}^-(x^{\mu}_-)$, in terms of independently defined
coordinate systems $x^{\mu}_+$ and $x^{\mu}_-$. The manifolds are
bounded by hypersurfaces $\Sigma_+$ and $\Sigma_-$, respectively,
with induced metrics $g_{ij}^+$ and $g_{ij}^-$. The hypersurfaces
are isometric, i.e., $g_{ij}^+(\xi)=g_{ij}^-(\xi)=g_{ij}(\xi)$, in
terms of the intrinsic coordinates, invariant under the isometry.
A single manifold ${\cal M}$ is obtained by gluing together ${\cal
M_+}$ and ${\cal M_-}$ at their boundaries, i.e., ${\cal M}={\cal
M_+}\cup {\cal M_-}$, with the natural identification of the
boundaries $\Sigma=\Sigma_+=\Sigma_-$.
In particular, assuming the continuity of the four-dimensional
coordinates $x^{\mu}_\pm$ across $\Sigma$, then $g_{\mu
\nu}^-=g_{\mu \nu}^+$ is required, which together with the
continuous derivatives of the metric components $\partial g_{\mu
\nu}/\partial x^\alpha|_-=\partial g_{\mu \nu}/\partial
x^\alpha|_+$, provide the Lichnerowicz conditions~\cite{Lich}.

The three holonomic basis vectors ${\bf e}_{(i)}=\partial
/\partial \xi^i$ tangent to $\Sigma$ have the following components
$e^{\mu}_{(i)}|_{\pm}=\partial x_{\pm}^{\mu}/\partial \xi^i$,
which provide the induced metric on the junction surface by the
following scalar product
\begin{equation}
g_{ij}={\bf e}_{(i)}\cdot {\bf e}_{(j)}=g_{\mu
\nu}e^{\mu}_{(i)}e^{\nu}_{(j)}|_{\pm}.
\end{equation}

We shall consider a timelike junction surface $\Sigma$, defined by
the parametric equation of the form $f(x^{\mu}(\xi^i))=0$. The
unit normal $4-$vector, $n^{\mu}$, to $\Sigma$ is defined as
\begin{equation}\label{defnormal}
n_{\mu}=\pm \,\left |g^{\alpha \beta}\,\frac{\partial f}{\partial
x ^{\alpha}} \, \frac{\partial f}{\partial x ^{\beta}}\right
|^{-1/2}\;\frac{\partial f}{\partial x^{\mu}}\,,
\end{equation}
with $n_{\mu}\,n^{\mu}=+1$ and $n_{\mu}e^{\mu}_{(i)}=0$. The
Israel formalism requires that the normals point from ${\cal M_-}$
to ${\cal M_+}$ \cite{Israel}.

The extrinsic curvature, or the second fundamental form, is
defined as $K_{ij}=n_{\mu;\nu}e^{\mu}_{(i)}e^{\nu}_{(j)}$, or
\begin{eqnarray}\label{extrinsiccurv}
K_{ij}^{\pm}=-n_{\mu} \left(\frac{\partial ^2 x^{\mu}}{\partial
\xi ^{i}\,\partial \xi ^{j}}+\Gamma ^{\mu \pm}_{\;\;\alpha
\beta}\;\frac{\partial x^{\alpha}}{\partial \xi ^{i}} \,
\frac{\partial x^{\beta}}{\partial \xi ^{j}} \right) \,.
\end{eqnarray}
Note that for the case of a thin shell $K_{ij}$ is not continuous
across $\Sigma$, so that for notational convenience, the
discontinuity in the second fundamental form is defined as
$\kappa_{ij}=K_{ij}^{+}-K_{ij}^{-}$. In particular, the condition
that $g_{ij}^-=g_{ij}^+$, together with the continuity of the
extrinsic curvatures across $\Sigma$, $K_{ij}^-=K_{ij}^+$, provide
the Darmois conditions \cite{Darmois}.

Now, the Lanczos equations follow from the Einstein equations for
the hypersurface, and are given by
\begin{equation}
S^{i}_{\;j}=-\frac{1}{8\pi}\,(\kappa ^{i}_{\;j}-\delta
^{i}_{\;j}\kappa ^{k}_{\;k})  \,,
\end{equation}
where $S^{i}_{\;j}$ is the surface stress-energy tensor on
$\Sigma$.

The first contracted Gauss-Kodazzi equation or the ``Hamiltonian"
constraint
\begin{eqnarray}
G_{\mu \nu}n^{\mu}n^{\nu}=\frac{1}{2}\,(K^2-K_{ij}K^{ij}-\,^3R)\,,
    \label{1Gauss}
\end{eqnarray}
with the Einstein equations provide the evolution identity
\begin{eqnarray}
S^{ij}\overline{K}_{ij}=-\left[T_{\mu \nu}n^{\mu}n^{\nu}
\right]^{+}_{-}\,.
\end{eqnarray}
The convention $\left[X \right]^+_-\equiv
X^+|_{\Sigma}-X^-|_{\Sigma}$ and $\overline{X} \equiv
(X^+|_{\Sigma}+X^-|_{\Sigma})/2$ is used.

The second contracted Gauss-Kodazzi equation or the ``ADM"
constraint
\begin{eqnarray}
G_{\mu \nu}e^{\mu}_{(i)}n^{\nu}=K^j_{i|j}-K,_{i}\,,
    \label{2Gauss}
\end{eqnarray}
with the Lanczos equations gives the conservation identity
\begin{eqnarray}\label{conservation}
S^{i}_{j|i}=\left[T_{\mu \nu}e^{\mu}_{(j)}n^{\nu}\right]^+_-\,.
\end{eqnarray}

In particular, considering spherical symmetry considerable
simplifications occur, namely $\kappa ^{i}_{\;j}={\rm diag}
\left(\kappa ^{\tau}_{\;\tau},\kappa ^{\theta}_{\;\theta},\kappa
^{\theta}_{\;\theta}\right)$. The surface stress-energy tensor may
be written in terms of the surface energy density, $\sigma$, and
the surface pressure, ${\cal P}$, as $S^{i}_{\;j}={\rm
diag}(-\sigma,{\cal P},{\cal P})$. The Lanczos equations then
reduce to
\begin{eqnarray}
\sigma &=&-\frac{1}{4\pi}\,\kappa ^{\theta}_{\;\theta} \,,\label{sigma} \\
{\cal P} &=&\frac{1}{8\pi}(\kappa ^{\tau}_{\;\tau}+\kappa
^{\theta}_{\;\theta}) \,. \label{surfacepressure}
\end{eqnarray}

\section{Generic dynamic spherically symmetric thin shells}

\subsection{Junction conditions}

We shall consider, in particular, the matching of two static and
spherically symmetric spacetimes given by the following line
elements
\begin{eqnarray}
ds^2_{\pm}&=&-e^{2\alpha_{\pm}(r_{\pm})}\,dt_{\pm}^2
+e^{2\beta_{\pm}(r_{\pm})}\,dr_{\pm}^2+r_{\pm}^2(d\theta_{\pm}
^2+\sin ^2{\theta_{\pm}}\, d\phi ^2_{\pm})  \,,
\label{generalmetric}
\end{eqnarray}
of ${\cal M_\pm}$, respectively. Using the Einstein field
equation, $G_{\hat{\mu}\hat{\nu}}=8\pi \,T_{\hat{\mu}\hat{\nu}}$,
in an orthonormal reference frame, (with $c=G=1$) the
stress-energy tensor components are given by
\begin{eqnarray}
\rho(r)&=&\frac{1}{8\pi} \;\frac{e^{-2\beta}}{r^2}
\,\left(2\beta'r+e^{2\beta}-1 \right)  \label{rho}\,,\\
p_r(r)&=&\frac{1}{8\pi} \frac{e^{-2\beta}}{r^2}
\,\left(2\alpha'r-e^{2\beta}+1 \right) \label{pr}\,,\\
p_t(r)&=&\frac{1}{8\pi}
\frac{e^{-2\beta}}{r}\,\left[-\beta'+\alpha'+r\alpha''
+r(\alpha')^2-r\alpha'
\beta'\right] \label{pt}\,,
\end{eqnarray}
where we have dropped the $\pm$ subscripts as not to overload the
notation, and in which $\rho(r)$ is the energy density, $p_r(r)$
is the radial pressure, and $p_t(r)$ is the lateral pressure
measured in the orthogonal direction to the radial direction.

The energy conditions will play an important role in the analysis
that follows, so we will at this stage define the null energy
condition (NEC). The latter is satisfied if
$T_{\mu\nu}\,k^\mu\,k^\nu \geq 0$, where $T_{\mu\nu}$ is the
stress-energy tensor and $k^{\mu}$ any null vector. Along the
radial direction, with $k^{\hat{\mu}}=(1,\pm 1,0,0)$ in the
orthonormal frame, we then have the following condition
\begin{equation}
T_{\hat{\mu}\hat{\nu}}\,k^{\hat{\mu}}\,k^{\hat{\nu}}=\rho(r)+p_r(r)=\frac{1}{4\pi
r}\, e^{-2\beta}\,(\alpha'+\beta') \geq 0  \,.  \label{generalNEC}
\end{equation}

The single manifold, ${\cal M}$, is obtained by gluing ${\cal
M_+}$ and ${\cal M_-}$ at $\Sigma$, i.e., at
$f(r,\tau)=r-a(\tau)=0$. In order for these line elements to be
continuous across the junction, we impose the following coordinate
transformations
\begin{eqnarray}
\fl t_+=\frac{e^{\alpha_-(a)}}{e^{\alpha_+(a)}} \;\;t_-  \,,
     \qquad
\frac{dr_+}{dr_-}\Big|_{r=a}=\frac{e^{\beta_{-}(a)}}{e^{\beta_+(a)}}
\,,
     \qquad
\theta_+=\theta_-    \quad \hbox{and} \quad \phi_+\;=\;\phi_- \,.
\end{eqnarray}
The intrinsic metric to $\Sigma$ is thus provided by
\begin{equation}
ds^2_{\Sigma}=-d\tau^2 + a(\tau)^2 \,(d\theta ^2+\sin
^2{\theta}\,d\phi^2)  \,.
\end{equation}

The imposition of spherical symmetry is sufficient to conclude
that there is no gravitational radiation, independently of the
behavior of the junction surface. The position of the junction
surface is given by
$x^{\mu}(\tau,\theta,\phi)=(t(\tau),a(\tau),\theta,\phi)$, and the
respective $4$-velocity is
\begin{eqnarray}
U^{\mu}_{\pm}=\left(e^{\beta_{\pm}(a)-\alpha_{\pm}(a)}\,\sqrt{e^{-2\beta_{\pm}(a)}+\dot{a}^2}\;,
\dot{a},0,0 \right)  \,,
\end{eqnarray}
where the overdot denotes a derivative with respect to $\tau$.

The unit normal to the junction surface may be determined by
equation (\ref{defnormal}) or by the contractions,
$U^{\mu}n_{\mu}=0$ and $n^{\mu}n_{\mu}=+1$, and is given by
\begin{eqnarray}
n^{\mu}_{\pm}=\left(e^{\beta_{\pm}(a)-\alpha_{\pm}(a)}\;\dot{a}
,\sqrt{e^{-2\beta_{\pm}(a)}+\dot{a}^2},0,0 \right) \label{normal}
\,,
\end{eqnarray}

Using equation (\ref{extrinsiccurv}), the non-trivial components
of the extrinsic curvature are given by
\begin{eqnarray}
K ^{\theta
\;\pm}_{\;\;\theta}&=&\frac{1}{a}\,\sqrt{e^{-2\beta_{\pm}}+\dot{a}^2}\;,
 \label{genKplustheta}\\
K ^{\tau \;\pm}_{\;\;\tau}&=&\frac{\alpha'_{\pm}
\left(e^{-2\beta_{\pm}}+\dot{a}^2
\right)+\ddot{a}+\beta'_{\pm}\,\dot{a}^2}{\sqrt{e^{-2\beta_{\pm}}+\dot{a}^2}}
\;. \label{genKminustautau}
\end{eqnarray}
%
%\subsection{Surface stresses}
%
The Einstein equations, eqs.
(\ref{sigma})-(\ref{surfacepressure}), with the extrinsic
curvatures, eqs. (\ref{genKplustheta})-(\ref{genKminustautau}),
then provide us with the following expressions for the surface
stresses
\begin{eqnarray}
\sigma&=&-\frac{1}{4\pi a} \left[\sqrt{e^{-2\beta}+\dot{a}^2}
\;\right]^+_-
    \label{gen-surfenergy2}   ,\\
{\cal P}&=&\frac{1}{8\pi a} \left[\frac{\left(1+a\alpha'\right)
\left(e^{-2\beta}+\dot{a}^2
\right)+a\ddot{a}+\beta'\,a\,\dot{a}^2}{\sqrt{e^{-2\beta}+\dot{a}^2}}
\; \right]^+_-
    \label{gen-surfpressure2}    \,.
\end{eqnarray}
If the surface stress-energy terms are zero, the junction is
denoted as a boundary surface. If surface stress terms are
present, the junction is called a thin shell. Note that the
surface mass of the thin shell is given by $m_s=4\pi a^2\sigma$

\subsection{Conservation identity}

Taking into account the transparency condition, $\left[G_{\mu
\nu}U^{\mu}\,n^{\nu}\right]^+_-=0$, the conservation identity,
equation (\ref{conservation}), provides the simple relationship
\begin{equation}
\dot{\sigma}=-2\frac{\dot{a}}{a}\,(\sigma +{\cal P})
    \,,
\end{equation}
or
\begin{equation}
\frac{d(\sigma A)}{d\tau}+{\cal P}\,\frac{dA}{d\tau}=0
    \,,
\end{equation}
where $A=4\pi a^2$ is the area of the thin shell. The first term
represents the variation of the internal energy of the shell, and
the second term is the work done by the shell's internal force.

Now, taking into account the momentum flux term in equation
(\ref{conservation}), $\left[T_{\mu
\nu}e^{\mu}_{(\tau)}n^{\nu}\right]^+_-=\left[T_{\mu
\nu}U^{\mu}\,n^{\nu}\right]^+_-$, we have
\begin{equation}
\left[T_{\mu
\nu}U^{\mu}\,n^{\nu}\right]^+_-=\left[\left(T_{\hat{t}\hat{t}}
+T_{\hat{r}\hat{r}}\right)e^{2\beta}\,\dot{a}\,\sqrt{e^{-2\beta}\,+\dot{a}^2}\;\right]^+_-
     \,,   \label{flux}
\end{equation}
where $T_{\hat{t}\hat{t}}$ and $T_{\hat{r}\hat{r}}$ are the
stress-energy tensor components given in an orthonormal basis. The
flux term, equation (\ref{flux}), corresponds to the net
discontinuity in the momentum flux $F_\mu=T_{\mu\nu}\,U^\nu$ which
impinges on the shell. Thus in the general case, the conservation
identity provides the following relationship
\begin{equation}
\sigma'=-\frac{2}{a}\,(\sigma +{\cal P})+\Xi
    \,,    \label{cons-equation}
\end{equation}
where for notational convenience, we have defined $\Xi$ as
\begin{equation}
\Xi=\left[\left(T_{\hat{t}\hat{t}}
+T_{\hat{r}\hat{r}}\right)e^{2\beta}\,\sqrt{e^{-2\beta}+\dot{a}^2}\;\right]^+_-
  =\frac{1}{4\pi a}\,
\left[\left(\alpha'+\beta'\right)\,\sqrt{e^{-2\beta}+\dot{a}^2}\;\right]^+_-
    \,.
\end{equation}
Note that this flux term vanishes in the particular case of
$p=-\rho$.

Taking into account the following relationship
\begin{equation}
\sigma+{\cal P}=\frac{1}{8\pi a}
\left[\frac{\left(a\alpha'-1\right)
e^{-2\beta}+a(\alpha'+\beta')\dot{a}^2-\dot{a}^2+a\ddot{a}}
{\sqrt{e^{-2\beta}+\dot{a}^2}} \; \right]^+_-   \,,
\end{equation}
and the definition of $\Xi$, equation (\ref{cons-equation}) takes
the form
\begin{equation}
\sigma'=\frac{1}{4\pi a^2}\,\left[\frac{\left(1+a\beta'\right)
\,e^{-2\beta}+\dot{a}^2-a\ddot{a}}{\sqrt{e^{-2\beta}+\dot{a}^2}}
\; \right]^+_-    \,,   \label{sigma'}
\end{equation}
which can also be obtained by taking the radial derivative of
equation (\ref{gen-surfenergy2}). Evaluated at the static solution
$a_0$, with $\dot{a}=\ddot{a}=0$, this reduces to
\begin{equation}
\sigma'(a_0)=\frac{1}{4\pi a_0^2}\left[e^{-\beta}\,(1+a_0\beta')
\right]^+_- \,.
                   \label{sigma_0'}
\end{equation}

Now, using $m_s=4\pi a^2 \sigma$, and taking into account the
radial derivative of $\sigma'$, equation (\ref{cons-equation}) can
be rearranged to provide the following relationship
\begin{equation}
\left(\frac{m_s}{2a}\right)''= \overline{\Upsilon}-4\pi
\sigma'\eta \,,
     \label{cons-equation2}
\end{equation}
with the parameter $\eta$ defined as $\eta={\cal P}'/\sigma'$, and
$\overline{\Upsilon}$ given by
\begin{equation}
\overline{\Upsilon}\equiv \frac{4\pi}{a}\,(\sigma+{\cal P})+2\pi a
\, \Xi '    \,.
    \label{Upsilon}
\end{equation}
The physical interpretation of $\eta$ is extensively discussed in
\cite{Poisson,Ishak}, and $\sqrt{\eta}$ is normally interpreted as
the speed of sound. Equation (\ref{cons-equation2}) will play a
fundamental role in determining the stability regions of the
respective solutions which will be analyzed further ahead.

For self-completeness, we shall also add the expression of $\Xi
'$, given by
\begin{equation}
\fl \Xi '=\frac{1}{4\pi
a^2}\,\left[[a(\alpha''+\beta'')-(\alpha'+\beta')]\,
\sqrt{e^{-2\beta}+\dot{a}^2}
+\frac{a(\alpha'+\beta')(-\beta'\,e^{-2\beta}+\ddot{a})}
{\sqrt{e^{-2\beta}+\dot{a}^2}}\right]^+_-
   \,,
\end{equation}
which evaluated at the static solution, $a_0$, reduces to
\begin{equation}
\Xi '_0=\frac{1}{4\pi a_0^2}\,\left[\left[a_0(\alpha''+\beta'')
-(\alpha'+\beta')(1+a_0\beta')\right]e^{-\beta}\right]^+_-
   \,,
\end{equation}

\subsection{Equation of motion}

Rearranging equation (\ref{gen-surfenergy2}) into the form
\begin{eqnarray}
\sqrt{e^{-2\beta_{+}}+\dot{a}^2}= \sqrt{e^{-2\beta_{-}}+\dot{a}^2}
- 4\pi a\,\sigma   \,,
\end{eqnarray}
we deduce the thin shell's equation of motion, i.e.,
\begin{equation}
\dot{a}^2+V(a)=0  \,.
\end{equation}
The potential $V(a)$ is given by
\begin{equation}
V(a)=F(a)-\left(\frac{m_s}{2a}\right)^2-\left(\frac{aG}{m_s}\right)^2
\,,
\end{equation}
where $m_s(a)=4\pi a^2\,\sigma$ is the mass of the thin shell. The
factors $F(a)$ and $G(a)$, introduced for computational
convenience, are defined by
\begin{eqnarray}
F(a)&\equiv&\overline{e^{-2\beta}}=\frac{1}{2}\,\left(e^{-2\beta_{-}}
+e^{-2\beta_{+}}\right)
\,,      \label{factorF}   \\
G(a)&\equiv&-\frac{1}{2}\left[e^{-2\beta}\right]^+_-=
\frac{1}{2}\,\left(e^{-2\beta_{-}}-e^{-2\beta_{+}}\right) \,,
         \label{factorG}
\end{eqnarray}
respectively.

Linearizing around a static solution situated at $a_0$, we
consider a Taylor expansion of $V(a)$ around $a_0$ to second
order, given by
\begin{equation}
V(a)=V(a_0)+V'(a_0)(a-a_0)+\frac{1}{2}V''(a_0)(a-a_0)^2+O[(a-a_0)^3]
\,.   \label{linear-potential}
\end{equation}
Note that one presumes that $V(a_0)=0$, which is imposed in order
to make the expansion consistent. The first and second derivatives
of $V(a)$ are, respectively, given by
\begin{eqnarray}
V'(a)=F'-2\left(\frac{m_s}{2a}\right)\left(\frac{m_s}{2a}\right)'
-2\left(\frac{aG}{m_s}\right)\left(\frac{aG}{m_s}\right)'
  \,,      \\
\fl V''(a)=F''-2\left[\left(\frac{m_s}{2a}\right)'\right]^2-
2\left(\frac{m_s}{2a}\right)\left(\frac{m_s}{2a}\right)''
-2\left[\left(\frac{aG}{m_s}\right)'\right]^2
-2\left(\frac{aG}{m_s}\right)\left(\frac{aG}{m_s}\right)'' \,.
\end{eqnarray}
Evaluated at the static solution, at $a=a_0$, through a long and
tedious calculation, we find $V(a_0)=0$ and $V'(a_0)=0$. Thus, the
potential, equation (\ref{linear-potential}), reduces to
$V(a)=\frac{1}{2}V''(a_0)(a-a_0)^2+O[(a-a_0)^3]$. From the
condition $V'(a_0)=0$, one extracts the following useful
equilibrium relationship
\begin{eqnarray}
\left(\frac{m_s}{2a_0}\right)'\equiv
\Gamma=\left(\frac{a_0}{m_s}\right)\left[
F'-2\left(\frac{a_0G}{m_s}\right)\left(\frac{a_0G}{m_s}\right)'\right]
  \,.
     \label{Gamma}
\end{eqnarray}

Note that if $V''(a_0)<0$ is verified, the potential $V(a)$ has a
local maximum at $a_0$, where a small perturbation in the surface
radius will produce an irreversible contraction or expansion of
the shell. Therefore, the solution is stable if and only if $V(a)$
has a local minimum at $a_0$ and $V''(a_0)>0$ is verified. The
latter stability condition takes the form
\begin{eqnarray}
\left(\frac{m_s}{2a}\right)\left(\frac{m_s}{2a}\right)''<\Psi
-\Gamma^2 \,,
     \label{masterequation}
\end{eqnarray}
where $\Psi$ is defined as
\begin{eqnarray}
\Psi=\frac{F''}{2}-\left[\left(\frac{aG}{m_s}\right)'\right]^2
-\left(\frac{aG}{m_s}\right)\left(\frac{aG}{m_s}\right)'' \,.
     \label{Psi}
\end{eqnarray}

\subsection{The master equation}

Substituting equation (\ref{cons-equation2}) into equation
(\ref{masterequation}), one deduces the master equation given by
\begin{equation}
\sigma' \,m_s \,\eta_0 > \Theta\,,
\end{equation}
where $\eta_0=\eta(a_0)$ and $\Theta$ is defined as
\begin{equation}
\Theta \equiv \frac{a_0}{2\pi} \left(\Gamma^2-\Psi \right)
+\frac{1}{4\pi}\,m_s\,\overline{\Upsilon}   \,.
       \label{master}
\end{equation}

Now, from the master equation we deduce that the stable
equilibrium regions are dictated by the following inequalities
\begin{eqnarray}
\eta_0 &>& \overline{\Theta}, \qquad {\rm if} \qquad \sigma'
\,m_s>0\,,      \label{stability1}
       \\
\eta_0 &<& \overline{\Theta}, \qquad {\rm if} \qquad \sigma'
\,m_s<0\,,       \label{stability2}
\end{eqnarray}
with the definition
\begin{eqnarray}
\overline{\Theta}\equiv \frac{\Theta}{\sigma' \,m_s}\,.
\end{eqnarray}
In the specific cases that follow, the explicit form of
$\overline{\Theta}$ may become extremely messy, so that as in
\cite{Ishak}, we find it more instructive to show the stability
regions graphically.

\section{A dynamic shell around a black hole}

Consider that the interior and exterior spacetimes are given by
equation (\ref{generalmetric}), with the following definitions
\begin{eqnarray}
e^{2\alpha_{\pm}}&=&e^{-2\beta_{\pm}}=\left(1-\frac{2m_{\pm}}{r_{\pm}}\right)
\label{metricvacuumlambda} \,.
\end{eqnarray}
with $m_+=M$ and $m_-=m$. These are separated by a thin shell with
the surface stresses given by
\begin{eqnarray}
\sigma&=&-\frac{1}{4\pi a} \left(\sqrt{1-\frac{2M}{a}+\dot{a}^2}-
\sqrt{1-\frac{2m}{a}+\dot{a}^2} \, \right)
    \label{BHsurfenergy}   ,\\
{\cal P}&=&\frac{1}{8\pi a}
\left(\frac{1-\frac{M}{a}+\dot{a}^2+a\ddot{a}}{\sqrt{1-\frac{2M}{a}+\dot{a}^2}}-
\frac{1-\frac{m}{a}+\dot{a}^2+a\ddot{a}}{\sqrt{1-\frac{2m}{a}+\dot{a}^2}}
\, \right)
    \label{BHsurfpressure}    ,
\end{eqnarray}
with the junction radius at $a>2M$.

For this case the transparency condition holds, as
$T_{\hat{t}\hat{t}}=T_{\hat{r}\hat{r}}=0$ and consequently
$\Xi=0$, so that the conservation identity provides the following
simple relationship
\begin{equation}
\sigma'=-\frac{2}{a}\,(\sigma +{\cal P})  \,.
     \label{EOS-BH}
\end{equation}
Taking into account equations
(\ref{BHsurfenergy})-(\ref{BHsurfpressure}), then equation
(\ref{EOS-BH}) takes the form
\begin{eqnarray}
\sigma'=\frac{1}{4\pi a^2} \left(\frac{1-\frac{3M}{a}
+\dot{a}^2-a\ddot{a}}{\sqrt{1-\frac{2M}{a}+\dot{a}^2}}-
\frac{1-\frac{3m}{a}
+\dot{a}^2-a\ddot{a}}{\sqrt{1-\frac{2m}{a}+\dot{a}^2}} \, \right)
\,,
     \label{sigma'BH}
\end{eqnarray}
and at the static solution $a_0$ reduces to
\begin{eqnarray}
\sigma'(a_0)&=&\frac{1}{4\pi a_0^2}
\left(\frac{1-\frac{3M}{a_0}}{\sqrt{1-\frac{2M}{a_0}}}-
\frac{1-\frac{3m}{a_0}}{\sqrt{1-\frac{2m}{a_0}}} \, \right)   \,,
\end{eqnarray}
which plays a fundamental role in determining the stability
regions.

Evaluated at the static solution, $a_0$, equations
(\ref{BHsurfenergy})-(\ref{BHsurfpressure}) reduce to
\begin{eqnarray}
\sigma&=&-\frac{1}{4\pi a_0} \left(\sqrt{1-\frac{2M}{a_0}}-
\sqrt{1-\frac{2m}{a_0}} \, \right)
    \label{staticBHsurfenergy}   ,\\
{\cal P}&=&\frac{1}{8\pi a_0}
\left(\frac{1-\frac{M}{a_0}}{\sqrt{1-\frac{2M}{a_0}}}-
\frac{1-\frac{m}{a_0}}{\sqrt{1-\frac{2m}{a_0}}} \, \right)
    \label{staticBHsurfpressure}    .
\end{eqnarray}
The weak energy condition (WEC) holds if $\sigma \geq 0$ and
$\sigma+{\cal P} \geq 0$ are satisfied, and by continuity implies
the null energy condition (NEC), $\sigma+{\cal P} \geq 0$. The
strong energy condition (SEC) is satisfied if $\sigma+{\cal P}\geq
0$ and $\sigma+2{\cal P}\geq 0$. Note that if $m<M$, then the
surface energy density is positive, $\sigma>0$, and it can also be
shown that the tangential surface pressure is also positive,
${\cal P}>0$. Thus, if $m<M$, the NEC, WEC and the SEC are readily
satisfied. If $m>M$, then the surface energy density and the
surface pressure are negative, consequently violating the energy
conditions.

The equation of motion is of the form $\dot{a}^2+V(a)=0$. Using
the factors of equations (\ref{factorF})-(\ref{factorG}), we have
\begin{equation}
F=1-\frac{m+M}{a}  \qquad  {\rm and}  \qquad G=\frac{M-m}{a} \,,
\end{equation}
and the potential is given by
\begin{equation}
V(a)=1-\frac{M+m}{a}-\frac{m_s^2}{4a^2}
-\frac{\left(M-m\right)^2}{m_s^2} \,.
\end{equation}

\begin{figure}[h]
\centering
  \includegraphics[width=2.6in]{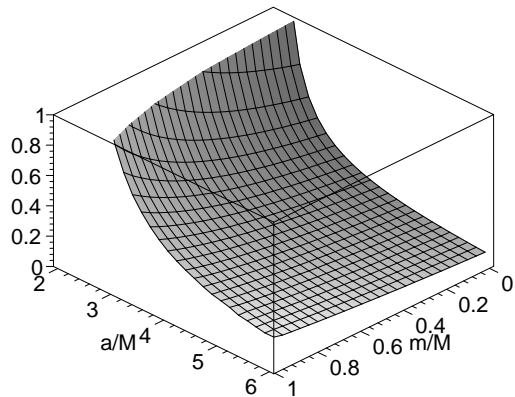}
  \caption{The regions of stability are situated below the
  surface, which is given by $\overline{\Theta}$.
  This is in agreement with previous results. See the text for details.}
  \label{BHstability}
\end{figure}

If $m<M$, then $m_s(a)>0$ and $\sigma'(a)<0$. If $m>M$, then
$m_s(a)<0$ and $\sigma'(a)>0$. For both cases the master equation
$\sigma'm_s\,\eta_0>\Theta$ is satisfied, so that the stability
regions supersede the energy conditions. Thus, equation
(\ref{master}), dictates the following regions of stability
\begin{equation}
\eta_0 < \overline{\Theta}  \,,
      \label{BHmasterequation}
\end{equation}
which is shown below the surface plotted in figure
\ref{BHstability}. This is in agreement with previous results
\cite{BLP,Ishak}.

\section{A dynamic shell around a traversable wormhole}

\subsection{Interior and exterior solutions}

Consider that the interior and exterior spacetimes are given by
equation (\ref{generalmetric}), with the following definitions
\begin{eqnarray}
e^{2\alpha_+}&=&e^{-2\beta_+}=\left(1-\frac{2M}{r_+}\right)  \,,
     \\
\alpha_-&=&\Phi(r_-)\,,  \qquad
e^{-2\beta_-}=\left(1-\frac{b(r_-)}{r_-}\right) \,. \label{defWH}
\end{eqnarray}
We have considered that the exterior spacetime is given by the
Schwarzschild solution. We shall once again drop the $(\pm)$
subscripts, as not to overload the notation. The spacetime with
the definition (\ref{defWH}) describes a wormhole geometry
\cite{Visser,Morris}, where $\Phi(r)$ and $b(r)$ are arbitrary
functions of the radial coordinate, $r$, denoted as the redshift
function and the form function, respectively \cite{Morris}. Note
that both solutions are matched at a junction interface, $r=a$,
situated outside the event horizon, i.e., $a>r_b=2M$, to avoid a
black hole solution.

Relatively to the interior solution, the field equations, eqs.
(\ref{rho})-(\ref{pt}), with the definition (\ref{defWH}) provide
the following stress-energy scenario
\begin{eqnarray}
\rho(r)&=&\frac{1}{8\pi} \,\frac{b'}{r^2}  \label{rhoWHlambda}\,,\\
p_r (r)&=&\frac{1}{8\pi} \left[-\frac{b}{r^3}+2
\left(1-\frac{b}{r}
\right) \frac{\Phi'}{r} \right]  \label{tauWHlambda}\,,\\
p_t(r)&=&\frac{1}{8\pi} \left(1-\frac{b}{r}\right)\Bigg[\Phi ''+
(\Phi')^2- \frac{b'r-b}{2r(r-b)}\Phi'
-\frac{b'r-b}{2r^2(r-b)}+\frac{\Phi'}{r} \Bigg]
\label{pressureWHlambda}\,,
\end{eqnarray}
in which $\rho(r)$ is the energy density; $p_r (r)$ is the radial
pressure; and $p(r)$ is the transverse pressure.

Wormhole spacetimes necessarily violate the null energy condition
(NEC) at the throat. The NEC as defined by equation
(\ref{generalNEC}) now takes the form
\begin{equation}\label{NECthroat}
T_{\hat{\mu}\hat{\nu}}k^{\hat{\mu}}k^{\hat{\nu}}=\rho(r)+p_r(r)=
\frac{1}{8\pi}\,\left[\frac{b'r-b}{r^3}+
2\left(1-\frac{b}{r}\right) \frac{\Phi '}{r} \right]  .
\end{equation}
Taking into account the flaring out condition of the throat
deduced from the mathematics of embedding, i.e., $(b-b'r)/2b^2>0$,
evaluated at the throat $b(r_0)=r=r_0$, and due to the finiteness
of $\Phi(r)$, we verify that equation (\ref{NECthroat}) is
necessarily negative, i.e.,
$T_{\hat{\mu}\hat{\nu}}k^{\hat{\mu}}k^{\hat{\nu}}<0$. Matter that
violates the NEC is denoted as {\it exotic matter}.

Recently, Visser, Kar and Dadhich, by implementing the notion of
the ``volume integral quantifier'', showed that the interior
wormhole solution may be supported by arbitrarily small quantities
of averaged null energy condition (ANEC) violating matter
\cite{VKD}, although the NEC and WEC are always violated for
wormhole spacetimes. In the spirit of minimizing the usage of the
exotic matter, regions where the surface stress-energy tensor
obeys the energy conditions at the junction were found
\cite{Lobo-CQG,Lobo}. Recently, it was also shown that traversable
wormholes may be supported by phantom energy
\cite{Sushkov,Lobo-phantom}, a null energy condition violating
cosmic fluid responsible for the present accelerated expansion of
the universe. In Ref. \cite{Lobo-phantom}, it was shown that these
phantom wormholes may be theoretically constructed by vanishing
amounts of ANEC violating phantom energy, and the stability
regions, in the spirit of this paper, were further explored in
Ref. \cite{dyn-phantomWH}.

\subsection{Stability regions}

The surface stresses, eqs.
(\ref{gen-surfenergy2})-(\ref{gen-surfpressure2}), for this
particular case takes the form
\begin{eqnarray}
\fl \sigma&=&-\frac{1}{4\pi a}
\left(\sqrt{1-\frac{2M}{a}+\dot{a}^2}-
\sqrt{1-\frac{b(a)}{a}+\dot{a}^2} \, \right)
    \label{surfenergy2}   ,\\
\fl {\cal P}&=&\frac{1}{8\pi a} \left(\frac{1-\frac{M}{a}
+\dot{a}^2+a\ddot{a}}{\sqrt{1-\frac{2M}{a}+\dot{a}^2}}-
\frac{(1+a\Phi') \left(1-\frac{b}{a}+\dot{a}^2
\right)+a\ddot{a}-\frac{\dot{a}^2(b-b'a)}{2(a-b)}}{\sqrt{1-\frac{b(a)}{a}+\dot{a}^2}}
\, \right)
    \label{surfpressure2}    .
\end{eqnarray}

The conservation identity provides us with
$\sigma'=-\frac{2}{a}\,(\sigma+{\cal P})+\Xi$, with $\Xi$ given by
\begin{eqnarray}
\Xi=-\frac{1}{8\pi}
 \frac{\sqrt{1-\frac{b(a)}{a}+\dot{a}^2}}{\left(1-\frac{b(a)}{a}\right)}
\left[\frac{b'(a)a-b(a)}{a^3}+2\left(1-\frac{b(a)}{a} \right)
\frac{\Phi'(a)}{a} \right]  \,.
       \label{H(a)}
\end{eqnarray}
Equation (\ref{sigma'}), for this particular case is given by
\begin{eqnarray}
\sigma'=\frac{1}{4\pi a^2} \left(\frac{1-\frac{3M}{a}
+\dot{a}^2-a\ddot{a}}{\sqrt{1-\frac{2M}{a}+\dot{a}^2}}-
\frac{1-\frac{3b(a)}{2a}+\frac{b'(a)}{2}+\dot{a}^2
-a\ddot{a}}{\sqrt{1-\frac{b(a)}{a}+\dot{a}^2}} \, \right) \,,
     \label{sigma'WH}
\end{eqnarray}
which at the static solution simplifies to
\begin{eqnarray}
\sigma'(a_0)&=&\frac{1}{4\pi a_0^2}
\left(\frac{1-\frac{3M}{a_0}}{\sqrt{1-\frac{2M}{a_0}}}-
\frac{1-\frac{3b(a_0)}{2a_0}+\frac{b'(a_0)}{2}}{\sqrt{1-\frac{b(a_0)}{a_0}}}
\, \right)   \,.
\end{eqnarray}
This relationship will play an important role in determining the
stability regions dictated by the master equation, equation
(\ref{master}).

The equation of motion is of the form $\dot{a}^2 + V(a)=0$, and
taking into account the factors of eqs.
(\ref{factorF})-(\ref{factorG}), which for this case are given by
\begin{equation}
F=1-\frac{b(a)/2+M}{a}  \qquad  {\rm and}  \qquad
G=\frac{M-b(a)/2}{a}  \,,
\end{equation}
the potential takes the form
\begin{equation}
V(a)=1-\frac{M+\frac{b(a)}{2}}{a}-\frac{m_s^2}{4a^2}
-\frac{\left(M-\frac{b(a)}{2}\right)^2}{m_s^2}
\end{equation}

Note that the functions $\Gamma$ and $\Psi$, given by eqs.
(\ref{Gamma}) and (\ref{Psi}), respectively, are completely
determined by the factors $F$, $G$ and $m_s=4\pi a^2 \sigma$. To
determine $\overline{\Upsilon}$, given by equation
(\ref{Upsilon}), and consequently $\Theta$, given by equation
(\ref{master}), one uses the radial derivative of equation
(\ref{H(a)}), and eqs. (\ref{surfenergy2})-(\ref{surfpressure2}),
evaluated at the static solution. One may now model the wormhole
geometry by choosing specific values for the form and redshift
functions, and consequently determine the stability regions
dictated by the inequalities
(\ref{stability1})-(\ref{stability2}). As the explicit form of
$\overline{\Theta}$ is extremely lengthy, so that as in Ref.
\cite{Ishak,dyn-phantomWH}, we find it more instructive to show
the stability regions graphically.

\subsection{Specific form and redshift functions}

We shall in this section consider various choices for the form
function, namely $b(r)=r_0$ and $b(r)=r_0^2/r$, and deduce the
respective stability regions. We shall verify that the latter may
be significantly increased by considering appropriate choices for
the redshift function. In this section, we shall relax the
condition that the surface energy density be positive, as in
considering traversable wormhole geometries, one is already
dealing with exotic matter.

\subsubsection{$b(r)=r_0$.}

Firstly, consider the particular case of $b(r)=r_0$ and
$\Phi'(a)=\Phi''(a)=0$. The redshift function can either be
constant, or have the following general form $\Phi(r)=\sum_{n\geq
3}\,c_n (r-a)^n$, so that the condition $\Phi'(a)=\Phi''(a)=0$ is
verified. In the analysis that follows below, only particular
cases shall be analyzed. Note that the factor $\Phi''(a)$ of the
model enters into $\Xi '$. As we are matching the interior
wormhole solution to an exterior vacuum spacetime, then there is
no need to impose the condition of an asymptotically flat
spacetime. For instance, this notion is reflected in the choice of
the redshift function given by $\Phi(r)=(r-a)^3/r_0^3$. However,
if one is tempted to model an asymptotically flat wormhole
geometry, in the absence of a thin shell, then one may use, for
instance, the choice of the redshift function given by
$\Phi(r)=r_0(r-a)^3/r^4$. In the analysis that follows we shall
separate the cases of $r_0<2M$ and $r_0>2M$, which corresponds to
a positive and negative surface energy densities.
\begin{figure}[h]
\centering
  \includegraphics[width=2.6in]{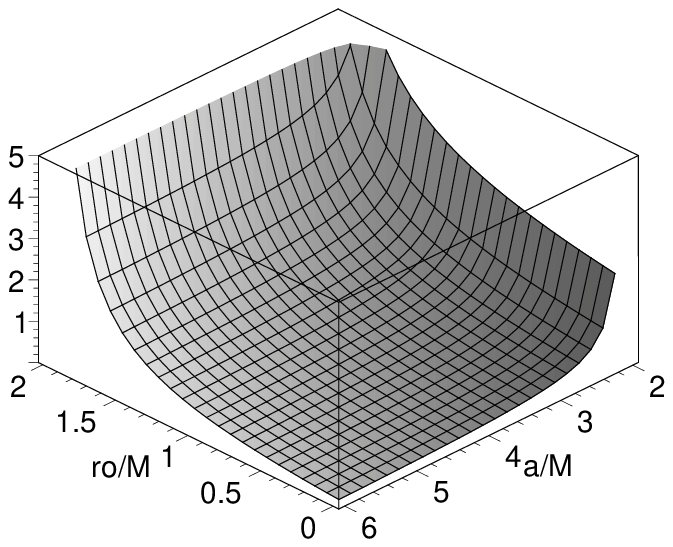}
  \hspace{0.4in}
  \includegraphics[width=2.6in]{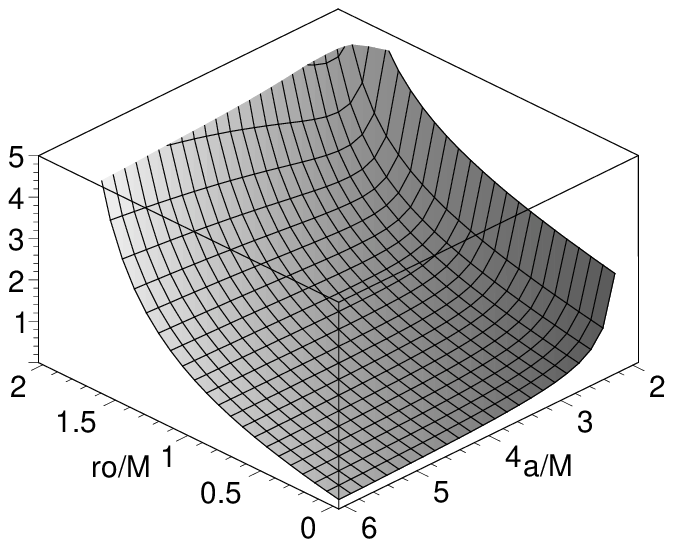}
  \caption{The stable equilibrium regions for the case of
  $r_0<2M$, with $b(r)=r_0$, are represented below
  the surfaces. We have considered specific choices for the
  redshift function, namely, $\Phi(r)=0$ and $\Phi(r)=r_0/r$,
  which are represented in the left and right plots,
  respectively. Note that, qualitatively,
  the stable equilibrium regions, for $\Phi(r)=r_0/r$, are
  increased relatively to the $\Phi(r)=0$ case.
  See the text for details.}\label{WHr0}
\end{figure}

For $r_0<2M$, we verify that $m_s(a_0)>0$ and $\sigma'(a_0)<0$,
and consequently $\sigma' m_s<0$, so that the stability regions
are dictated by inequality (\ref{stability2}). The stable
equilibrium configurations are shown below the surfaces of the
plots depicted in figure \ref{WHr0}. In the left plot, we have
considered the case of $\Phi(r)=0$. Now, it is interesting to note
that one may increase the stability regions by adequately choosing
a specific redshift function. To illustrate this, consider for
simplicity the choice of $\Phi(r)=r_0/r$. Note that qualitatively
the stable equilibrium regions, for this case, are increased for
high values of $a/M$ as $r_0 \rightarrow 2M$, relatively to the
$\Phi(r)=0$ case. It is possible to consider more complicated
cases, however, this shall not be done here.

\begin{figure}[h]
\centering
  \includegraphics[width=2.6in]{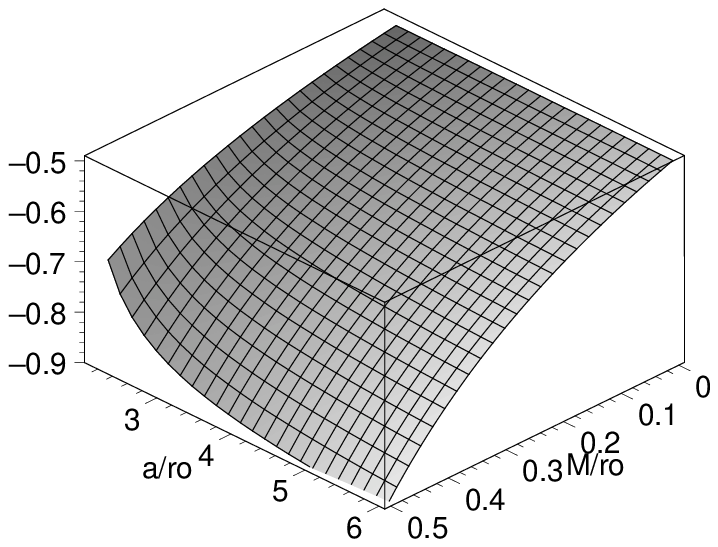}
  \hspace{0.4in}
  \includegraphics[width=2.6in]{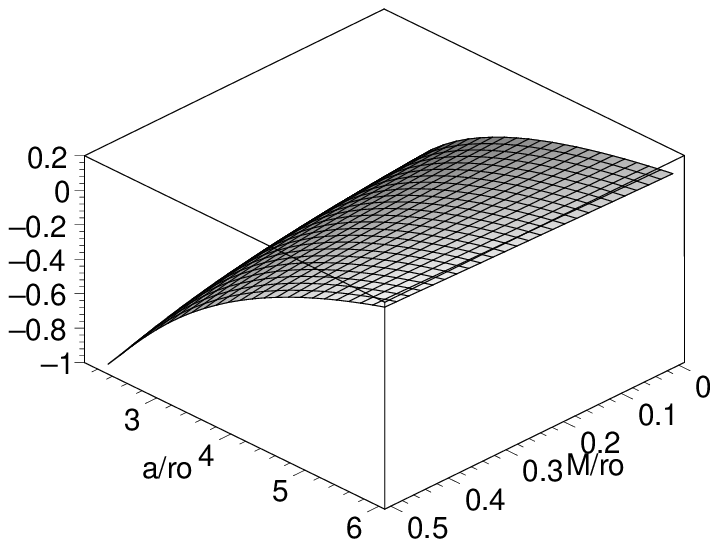}
  \caption{The stable equilibrium regions for the case of
  $r_0>2M$, with $b(r)=r_0$, are represented below
  the surfaces. We have considered specific choices for the
  redshift function, namely, $\Phi'(a)=\Phi''(a)=0$ and
  $\Phi(r)=-r_0/r$, which are represented on the left and
  right plots, respectively. The stability regions are
significantly increased, for the $\Phi(r)=-r_0/r$ case, as $r_0
\rightarrow 2M$. However, the stability regions for the
$\Phi'(a)=\Phi''(a)=0$ are greater for low values of $M/r_0$ and
$a/r_0$.}
  \label{WHr02}
\end{figure}

For $r_0>2M$, we verify that $m_s(a_0)<0$ and $\sigma'(a_0)>0$,
and the stable equilibrium configurations are also dictated by
inequality (\ref{stability2}). We have considered the specific
cases of $\Phi'(a)=\Phi''(a)=0$ and $\Phi(r)=-r_0/r$, represented
in the left and right plots of figure \ref{WHr02}, respectively.
For both cases the stability regions are also given below the
surfaces represented in the plots. Note that the stability regions
are significantly increased, for the $\Phi(r)=-r_0/r$ case, as
$r_0 \rightarrow 2M$, as can be readily verified from figure
\ref{WHr02}. However, the stability regions for the
$\Phi'(a)=\Phi''(a)=0$ are considerably greater for low values of
$M/r_0$ and of $a/r_0$. Thus, the message that one can extract
from this analysis is that one may model stable wormhole
geometries, by adequately choosing the redshift function.

\subsubsection{$b(r)=r_0^2/r$.}

Consider the Ellis drainhole \cite{ellis}, with $b(r)=r_0^2/r$ and
$\Phi(r)=0$. However as in the previous case, this analysis can be
extended to an arbitrary redshift function in which
$\Phi'(a)=\Phi''(a)=0$. Recently, in an interesting paper, it was
shown that this geometry can also be obtained with `tachyon
matter' as a source term in the field equations and a positive
cosmological constant \cite{Das-Kar}.

To determine the stability regions of this solution, we shall
separate the cases of $b(a_0)<2M$ and $b(a_0)>2M$. From equation
(\ref{surfenergy2}) and the definition of $m_s=4\pi a_0^2 \sigma$,
this corresponds to $m_s >0$ and $m_s <0$, respectively.

\begin{figure}[h]
\centering
  \includegraphics[width=2.6in]{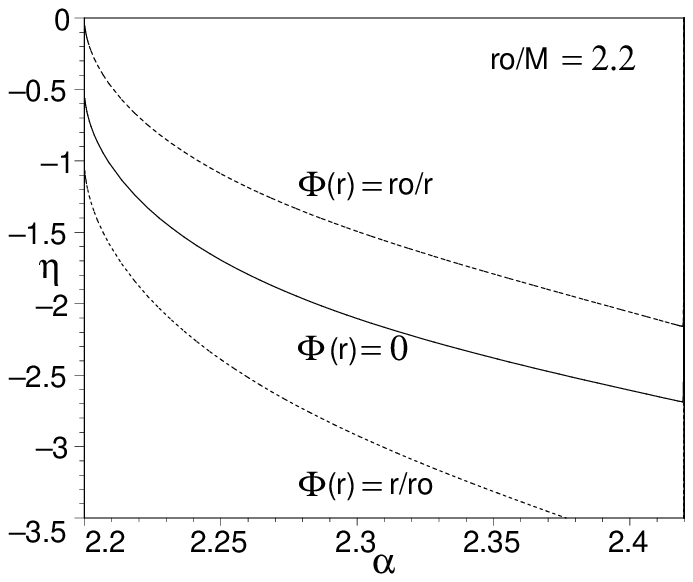}
  \hspace{0.4in}
  \includegraphics[width=2.6in]{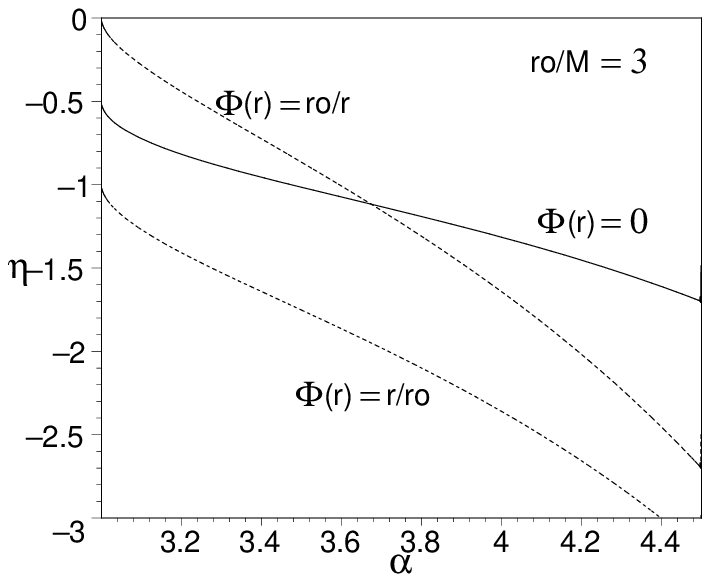}
  \caption{We have defined $\alpha=a_0/M$ and considered the
  specific cases of $r_0/M=2.2$ and $r_0/M=3$.
  The stability regions are depicted below the curves. For low
  values of $r_0/M$, we verify that considering the choice of
  $\Phi(r)=r_0/r$, the stability regions increase relatively to
  $\Phi(r)=0$. For instructive purposes, we have included the case
  of $\Phi(r)=r/r_0$, for which the stability regions decrease.
  See the text for details.}
  \label{WHdrain}
\end{figure}

For $b(a_0)>2M$, i.e., for a negative surface energy density, and
using the form function considered above, we need to impose the
condition $r_0>2M$, so that the junction radius lies outside the
event horizon, $a_0>2M$. Thus, the junction radius lies in the
following range
\begin{equation}
r_0< a_0 <\frac{r_0^2}{2M}  \,.
     \label{flat-range}
\end{equation}
We verify that $m_s \,\sigma'<0$, so that according to the
inequality (\ref{stability2}), the stability regions lie below the
curves depicted in figure \ref{WHdrain}. We have considered the
cases of $r_0/M=2.2$ and $r_0/M=3$, where the ranges of $a_0$ are
given by $2.2<a_0/M<2.42$ and $3<a_0/M<4.5$, respectively. Note
that the values of $\eta_0$ are negative in the stability regions,
however, by increasing the parameter $r_0/M$, for $\Phi(r)=0$, the
values of $\eta_0$ become less restricted, so that the stability
regions are increased.

Now, considering a fixed value of $r_0/M$ and for $\Phi(r)=0$,
considering the choice of $\Phi(r)=r_0/r$, we have certain regions
for which the stability increases. We have also included the
specific choice of $\Phi(r)=r/r_0$, for instructive purposes,
representing a specific case for which the stable equilibrium
configurations decrease.

For $b(a_0)<2M$, i.e., for a positive surface energy density, and
using the form function considered above, we shall separate the
cases of $r_0<2M$ and $r_0>2M$.

\begin{figure}[h]
\centering
  \includegraphics[width=2.6in]{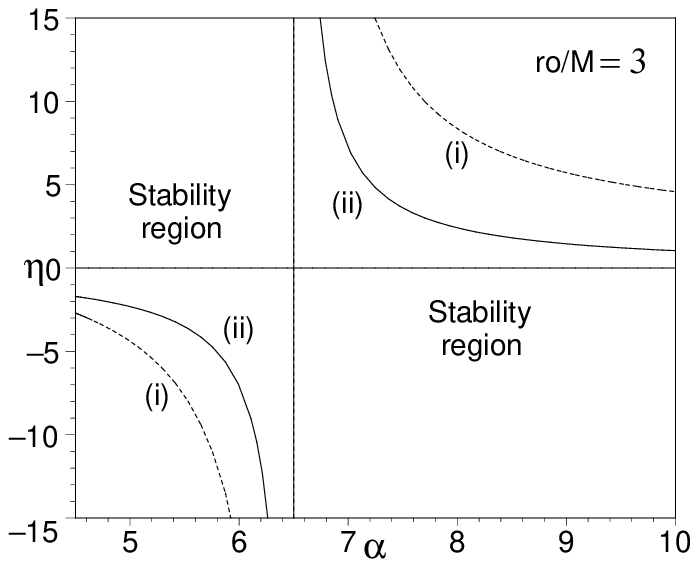}
  \hspace{0.4in}
  \includegraphics[width=2.6in]{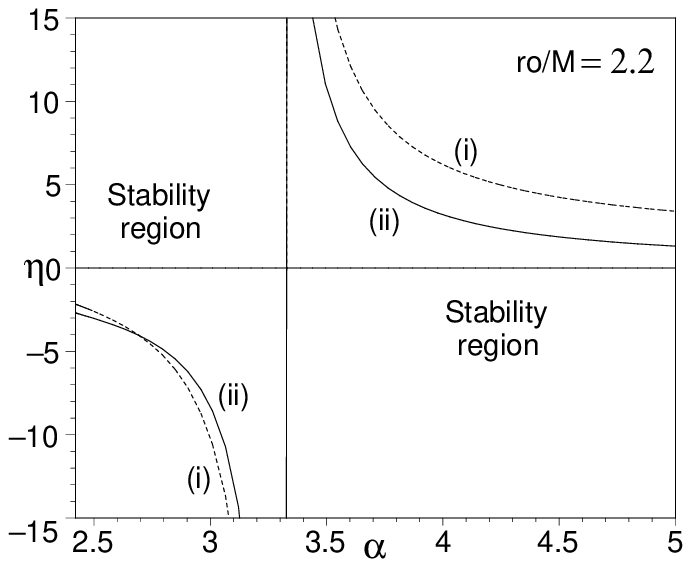}
  \caption{We have defined $\alpha=a_0/M$, and considered the
  cases of $r_0/M=3$ and $r_0/M=2.2$.
  The stability regions are depicted in the plots.
  $(i)$ corresponds to $\Phi(r)=r_0/r$; $(ii)$ corresponds to
  $\Phi(r)=0$. Note that one may increase the stability regions by
  choosing an appropriate redshift function. See the text for
  details.}
  \label{WHdrain2}
\end{figure}

Firstly, considering the case of $r_0>2M$, we verify that the
junction radius lies in the following range
\begin{equation}
a_0 >\frac{r_0^2}{2M}  \,.\label{range2}
\end{equation}
For this specific case, $\sigma'$ possesses one real positive
root, $R$, in the range of equation (\ref{range2}), signalling the
presence of an asymptote, $\sigma'|_R=0$. We verify that
$\sigma'>0$ for $r_0^2/(2M)<a_0 <R$, and $\sigma'<0$ for $a_0
>R$. Thus, the stability regions are given by
\begin{eqnarray}
\eta_0 &>& \overline{\Theta}, \quad {\rm if} \quad
\frac{r_0^2}{2M}<a_0 <R  \,,
         \label{drain-stability1}
       \\
\eta_0 &<& \overline{\Theta}, \quad {\rm if} \quad a_0 >R\,.
        \label{drain-stability2}
\end{eqnarray}
Consider the particular cases of $r_0/M=2.2$, so that
$a_0/M>2.42$, and $r_0/M=3$, where $a_0/M>4.5$. The asymptotes,
$\sigma'|_R=0$, for these cases exist at $R/M \simeq 3.285$ and
$R/M\simeq 6.50265$, respectively. These cases are represented in
figure \ref{WHdrain2}. We verify that, for $\Phi(r)=0$, and
considering a fixed high value of $r_0/M$, the specific choice of
$\Phi(r)=r_0/r$ increases the regions of stability. However, for
low values of the parameter $r_0/M$, and for low $a_0/M$, we need
to be more careful, as certain regions for the specific case of
$\Phi(r)=r_0/r$, decrease the stable equilibrium configurations of
the solution, as is qualitatively represented in the right plot of
figure \ref{WHdrain2}.

\begin{figure}[h]
\centering
  \includegraphics[width=2.6in]{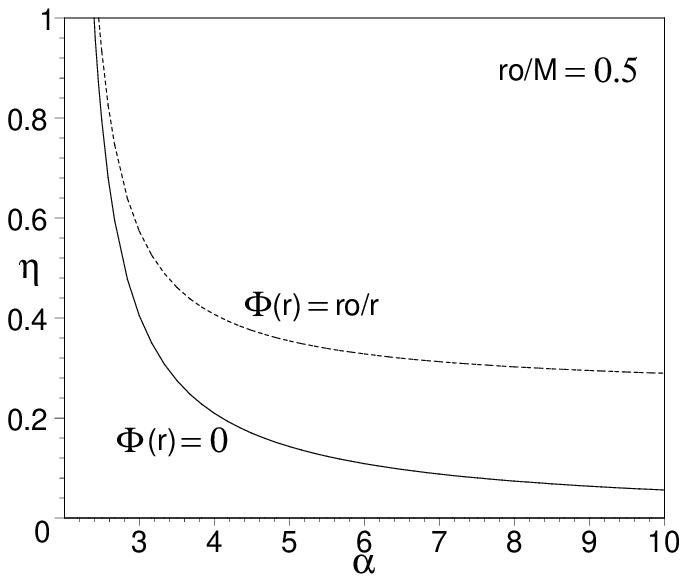}
  \hspace{0.4in}
  \includegraphics[width=2.6in]{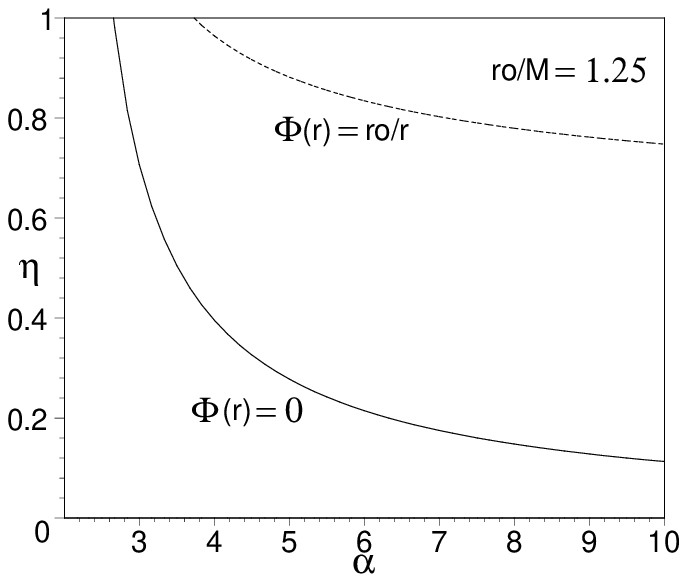}
  \caption{We have defined $\alpha=a_0/M$, and considered the
  specific choices of $r_0/M=0.5$ and $r_=/M=1.25$. The stability
  regions are depicted below the curves. Note that the stability
  regions for $\Phi(r)=0$ are increased by increasing the
  parameter $r_0/M$, and further significantly increased by
  choosing an appropriate redshift function, for instance,
  $\Phi(r)=r_0/r$. See the text for details.}
  \label{WHdrain3}
\end{figure}

Considering the case of $r_0<2M$, we verify that the junction
radius lies in the range $a_0>2M$. We have $m_s\,\sigma'<0$, so
that the stability region lies below the curves depicted in figure
\ref{WHdrain3}. Adopting a conservative point of view, we verify
that for this case of a positive surface energy density, stability
regions exist well within the range $0<\eta_0 \leq 1$. The
stability regions are increased by increasing the value of $
r_0/M$, with $\Phi(r)=0$, and is further increased by considering
the specific choice of $\Phi(r)=r_0/r$.

\section{Conclusion}

Thin shells play an extremely fundamental role in general
relativity, and have numerous applications ranging from the study
of gravitational collapse to the Randall-Sundrum brane world
scenario \cite{examples}. In this paper, we have analyzed the
stability of generic spherically symmetric thin shells to
linearized perturbations around static solutions. We have included
the momentum flux term in the conservation identity, deduced from
the ``ADM'' constraint and the Lanczos equations. Following the
Ishak-Lake analysis, a master equation dictating the stable
equilibrium configurations was deduced. Considering the
transparency condition, we studied the stability of thin shells
around black holes, showing that our analysis is in agreement with
previous results. Applying the analysis to traversable wormhole
geometries, considering specific choices for the form function, we
deduced stability regions, and found that the latter may be
significantly increased by considering appropriate choices for the
redshift function.

We have analyzed the stability by defining a parameter $\eta$, so
that one does not have to define an equation of state of the
stresses involved. Normally, $\sqrt{\eta}$ is interpreted as the
speed of sound, so that the requirement that the latter does not
exceed the speed of light is naturally $0\leq \eta<1$. Although we
have imposed this condition for the analysis concerning thin
shells around black holes, this definition cannot be naively
applied to stresses that violate the null energy condition, i.e.,
``exotic matter''. Therefore, we have relaxed the range $0\leq
\eta<1$, when considering the stability analysis for thin shells
around traversable wormholes, although stability regions do exist
for the referred interval.
It is also interesting to note, as emphasized in Ref.
\cite{Lobo-CQG}, that thin shells (or domain walls) arise in
models with spontaneously broken discrete symmetries in field
theory. Note that these models involve a set of real scalar fields
$\phi_i$, with a Lagrangian of the form ${\cal
L}=\frac{1}{2}\left(\partial_\mu \phi_i\right)^2-V(\phi)$, where
the potential $V(\phi)$ has a discrete set of degenerate minima.
Thus, by suitably choosing $\phi_i$ and $V(\phi)$, one may obtain
the dynamically stable thin shells analyzed in this work.

%%%%%%%%%%%%%%%%%%%%%%%%%%%%%%%%%%%%%%%%%%%%%%%%%%%%%%%%%%%%%%%%%%%
%%%%%%%%%%%%%%%%%%%%%%%%%%%%%%%%%%%%%%%%%%%%%%%%%%%%%%%%%%%%%%%%%%%%

\end{document}